# Photothermal Heterodyne Imaging of Individual Metallic Nanoparticles: Theory versus Experiments

Stéphane Berciaud, David Lasne, Gerhard A. Blab, Laurent Cognet & Brahim Lounis

*Centre de Physique Moléculaire Optique et Hertzienne, CNRS (UMR 5798) et Université Bordeaux I, 351, cours de la Libération, 33405 Talence Cedex, France*

*We present the theoretical and detailed experimental characterizations of Photothermal Heterodyne Imaging. An analytical expression of the photothermal heterodyne signal is derived using the theory of light scattering from a fluctuating medium. The amplitudes of the signals detected in the backward and forward configurations are compared and their frequency dependences are studied. The application of the Photothermal Heterodyne detection technique to the absorption spectroscopy of individual gold nanoparticles is discussed and the detection of small individual silver nanoparticles is demonstrated.*

# I Introduction

Several optical schemes have been used to perform the detection of nano-objects at the single entity level. Together with constant improvements in synthesis and characterization of nanosized materials, those highly sensitive methods make possible the development of new nano-components such as plasmonic devices[1,2] or single photon sources [3]. Until recently most optical detection methods of single nanometer sized-objects were based on luminescence. Single fluorescent molecules or semiconductor nanocrystals have been extensively studied and are now widely implemented in various research domains ranging from quantum optics[4] to life science[5]. However, luminescence based techniques suffer for some shortcomings, mainly associated with the photostability of the luminescent nano-object itself. Concurrently, for relatively large nanoparticles, Rayleigh scattering based methods have recently demonstrated a great applicability, especially for single metal nanoparticles (NPs) spectroscopy[6,7] or biomolecules imaging[8]. However, as the scattering cross-sections of the particles decrease with the sixth power of the diameter, these methods are limited to the study of rather large NPs (diameter > 20 nm).

Since the absorption cross-section of these NPs scales with the volume, an interesting alternative to Rayleigh-scattering relies solely on absorption properties. Indeed, excited near their plasmon resonance, metal NPs have a relatively large absorption cross section (~$6\times10^{-14}$ $cm^2$ for a 5 nm diameter gold NP) and exhibit a fast electron-phonon relaxation time in the picosecond range[9], which makes them very efficient light absorbers. The luminescence yield of these particles being extremely weak[10,11], almost all the absorbed energy is converted into heat. The increase of temperature induced by this absorption gives rise to a local variation of the refraction index. This photothermal effect was first used to detect gold NPs as small as 5 nm in diameter by a Photothermal Interference Contrast (PIC) method[12]. In that case, the signal was caused by the phase-shift between the two orthogonally polarized, spatially-separated beams of an interferometer, only one of which propagating through the heated region. The sensitivity of this technique, though high, is limited. In particular, when high NA objectives are used, depolarization effects degrade the quality of the overlap between the two arms of the interferometer.

We recently developed another photothermal method, called Photothermal Heterodyne Imaging (PHI)[13]. It uses a combination of a time-modulated heating beam and a non-resonant probe beam. The probe beam produces a frequency shifted scattered field as it interacts with the time modulated variations of the refractive index around the absorbing NP. The scattered field is then detected through its beatnote with the probe field which plays the role of a local oscillator as in any heterodyne technique. This signal is extracted by lock-in detection. The sensitivity of PHI lies two orders of magnitude above earlier methods[12] and it allowed for the unprecedented detection of individual 1.4 *nm* (67 atoms) gold NPs, as well as CdSe/ZnS semiconductor nanocrystals. In addition, since the PHI signal is directly proportional to the power absorbed by the nano-object, this method could be used to perform absorption spectroscopy studies of individual gold NPs down to diameters of 5 *nm*[14]

The goal of this paper is to give the theoretical framework of the PHI method and to compare the expected signals with the experimental results. In the following section, after a qualitative description of the principle of the PHI method, we will present an analytical derivation of the PHI signal. For this purpose the theory of light scattering by a fluctuating dielectric medium is used as in our case, the photothermal effect occurs on dimensions much smaller than the optical wavelength of the probe beam and previous derivations of photothermal techniques do not apply. The experimental study of the PHI signal is presented in Section III. Its variations with the modulation frequency are detailed both theoretically and experimentally in section

IV. In section V, we present the results of the implementation of PHI spectroscopy of individual gold NPs followed by preliminary results obtained with silver NPs. In the last section, we briefly discuss further implementations of PHI.

## II Theoretical model for the PHI signal

Throughout this article, we consider an absorbing nanosphere with radius $a$ much smaller than the optical wavelengths embedded in a homogeneous medium whose thermal diffusivity is $D = \kappa/C$ (with $\kappa$ the thermal conductivity of the medium and $C$ its heat capacity per unit volume). When illuminated with an intensity modulated laser beam with average intensity $I_{heat}$, the NP absorbs $P_{abs}[1 + \cos(\Omega t)]$, where $\Omega$ is the modulation frequency, $P_{abs} = \sigma\, I_{heat}$ with $\sigma$ the absorption cross section of the particle. At distance $\rho$ from the center of the NP (Fig. 1), the temperature rise can be derived using the heat point source model for heat diffusion[15]:

$$\Delta T(\rho,t) = \frac{P_{abs}}{4\pi\kappa\,\rho}\left[1 + \exp\left(-\frac{\rho}{r_{th}}\right)\cos\left(\Omega t - \frac{\rho}{r_{th}}\right)\right] \qquad (1)$$

with $r_{th} = \sqrt{\frac{2D}{\Omega}}$ the characteristic length for heat diffusion. The corresponding index of refraction profile is $\Delta n = \frac{\partial n}{\partial T}\Delta T$, where $\frac{\partial n}{\partial T}$ are the variations of the refractive index with temperature (typically, $10^{-4}\,K^{-1}$). In the following, we will rather use the induced susceptibility profile: $\Delta\chi = 2n\frac{\partial n}{\partial T}\Delta T$, where we denote $n$ the non-perturbed refractive index of the medium.

The geometry of the problem is represented in Figure 1. A focused circularly polarized probe beam interacts with the time modulated susceptibility profile. At the beam waist, a plane wave-front is considered. Thus, the incident probe field can be written $\mathbf{E}_i(\boldsymbol{\rho},t) = E_0 e^{i(\mathbf{k}_i \cdot \boldsymbol{\rho} - \omega t)} \mathbf{e}_+$ (with $\mathbf{e}_+ = \frac{\mathbf{e}_x + i\mathbf{e}_y}{\sqrt{2}}$ and $\omega$ the frequency of the incident field).

In practice, $r_{th}$ is smaller than the probe beam's wavelength (see below). Hence, a full electromagnetic derivation of the scattering field is necessary to further evaluate the detected signal. We used the model introduced by Lastovka[16] as a starting point for our derivation[17]. First, the interaction of the incident field with the local susceptibility fluctuations give rise to a polarization $\widetilde{\mathbf{P}}(\boldsymbol{\rho},t)$:

$$\widetilde{\mathbf{P}}(\boldsymbol{\rho},t) = \frac{\varepsilon_0 \Delta\chi(\rho,t)}{n^2} \mathbf{E}_i(\boldsymbol{\rho},t) \qquad (2)$$

The expression of the scattered field at a point M (with $\mathbf{OM} = \mathbf{R}$) is derived by introducing the Hertz potential[18] $\boldsymbol{\Pi}(\mathbf{R},t)$ which obeys to an inhomogeneous wave equation with the local polarization variations $\widetilde{\mathbf{P}}(\boldsymbol{\rho},t)$ as a source term:

$$\boldsymbol{\Pi}(\mathbf{R},t) = \frac{1}{4\pi\varepsilon_0} \int d^3\boldsymbol{\rho} \int dt' \frac{\widetilde{\mathbf{P}}(\boldsymbol{\rho},t')\delta(t'-t+|\mathbf{R}-\boldsymbol{\rho}|/c_m)}{|\mathbf{R}-\boldsymbol{\rho}|} \qquad (3)$$

$c_m$ being the speed of light in the medium. As the polarization variations $\widetilde{\mathbf{P}}$ are localized in the vicinity of the particle, at the observation point M, the total electric field is simply related to this Hertz potential by:

$$\mathbf{E}(\mathbf{R},t) = \nabla \times (\nabla \times \boldsymbol{\Pi}(\mathbf{R},t)) \qquad (4)$$

At this point, it should be stressed that only components $\mathbf{E}_\Omega(\mathbf{R},t)$ of the electric field which are frequency shifted by $\Omega$ from the incident field frequency will contribute to the PHI signal. Moreover, in the far-field, an analytical expression of $\mathbf{E}_\Omega(\mathbf{R},t)$ can be derived. Indeed, considering that the spatial extension of $\Delta\chi(\rho,t)$ is microscopic (of the order of $r_{th}$ and thus small compared to $|\mathbf{R}-\boldsymbol{\rho}|$), points M in the far-field region will be such that $r = |\mathbf{R}-\boldsymbol{\rho}| \approx R - \boldsymbol{\rho}.\mathbf{e}_R$. The zeroth-order of this expression will be taken in the denominator of the integrand of Eq. (3). Furthermore, considering that $c_m \gg r_{th}\Omega$, retardation effects can be neglected in the temporal integral of Eq. (3). Then, using Eq. (4) and introducing the vector $\Delta\mathbf{k} = \mathbf{k}_i - \mathbf{k}_s = \frac{2\pi n}{\lambda}(\mathbf{e}_z - \mathbf{e}_R)$, the electric field scattered at frequency shifted by $\pm\Omega$ with respect to incident field frequency $\omega$ writes in the far-field region:

$$\mathbf{E}_\Omega(\mathbf{R},t) = \frac{1}{2}\left(\mathbf{E}_\Omega^+(\mathbf{R},t) + \mathbf{E}_\Omega^-(\mathbf{R},t)\right) \tag{5}$$

with:

$$\mathbf{E}_\Omega^\pm(\mathbf{R},t) = 2n\frac{\partial n}{\partial T}\frac{P_{abs}}{(4\pi)^2\kappa}\frac{\varepsilon_0}{n^2}\frac{1}{R}\nabla\times\left[\nabla\times\mathbf{e}_+ I^\pm(\theta,\Omega)E_0 e^{i(k_s R-(\omega\pm\Omega)t)}\right] \tag{6}$$

and

$$I^\pm(\theta,\Omega) = \int \exp\left(-\frac{\rho}{r_{th}} + i\left(\Delta\mathbf{k}.\boldsymbol{\rho} \pm \frac{\rho}{r_{th}}\right)\right)\frac{d^3\boldsymbol{\rho}}{\rho} \tag{7}$$

After integration, one obtains:

$$I^\pm(\theta,\Omega) = 2\pi\, r_{th}^2\left(f(\theta,\Omega) \pm ig(\theta,\Omega)\right) \tag{8}$$

The functions $f$ and $g$ write after the introduction of the parameter $u(\theta,\Omega)=|\Delta k| \; r_{th} = \dfrac{4\pi n}{\lambda}\sqrt{\dfrac{2D}{\Omega}}\sin(\theta/2)$:

$$f(u) = \dfrac{1}{u}\left[\dfrac{u+1}{(u+1)^2+1} + \dfrac{u-1}{(u-1)^2+1}\right]$$
$$g(u) = \dfrac{1}{u}\left[\dfrac{1}{(u+1)^2+1} + \dfrac{-1}{(u-1)^2+1}\right] \quad (9)$$

The far-field component of the scattered field (which varies as $1/R$) has the form:

$$\mathbf{E}_\Omega(\mathbf{R},t) = -2\pi \, n \dfrac{\partial n}{\partial T}\dfrac{P_{abs}}{C\lambda^2}\left[\dfrac{f(\theta,\Omega)\cos(\Omega t)+g(\theta,\Omega)\sin(\Omega t)}{\Omega}\right]\dfrac{E_0}{R}e^{i(k_s R-\omega t)}\mathbf{e}_R \times (\mathbf{e}_R \times \mathbf{e}_+) \quad (10)$$

The detected signal will now be evaluated from this expression. It originates from the interference between $\mathbf{E}_\Omega(\mathbf{R},t)$ and a local oscillator field proportional to the incident probe field. Assuming that the wave front of this local oscillator is spherical in the far-field region, the power distribution of the resulting beatnote per unit of solid angle in the direction $(\theta,\phi)$ writes:

$$\dfrac{d^2 P_{PHI}(\theta,\Omega)}{\sin\theta \, d\theta \, d\phi} = \sqrt{2}\; n\dfrac{\partial n}{\partial T}\dfrac{P_{abs}}{C\lambda^2 w}\sqrt{P_i P_{LO}}\left[\dfrac{f(\theta,\Omega)\cos(\Omega t)+g(\theta,\Omega)\sin(\Omega t)}{\Omega}\right][1+|\cos\theta|] \quad (11)$$

with $P_i$ the incident power of the probe, $w$ the waist of the probe beam in the sample plane and $P_{LO}$ the power of the local oscillator. Experimentally we deal with two configurations where we detect either the backward or forward contribution of the scattered field. In the backward configuration, $P_{LO}$ is the reflected probe power at the interface between the glass slide and the medium surrounding the NPs. In the forward configuration, it is the transmitted probe power through the sample. More precisely, $P_{LO} = \alpha_{B/F} P_i$, where $\alpha_B = R$ and $\alpha_F = T$ are the intensity reflection and transmission coefficients at the glass/sample interface.

Integration of Eq. (11) leads to the beatnote power arriving on the detector and oscillating at frequency $\Omega$. In the backward/forward configuration, it writes:

$$P_{B/F}(\Omega,t) = \eta_{B/F} \sqrt{\alpha_{B/F}} P_i \left[ 2\pi\sqrt{2}.n \frac{\partial n}{\partial T} \frac{P_{abs}}{C\lambda^2 w} \right] \left[ F_{B/F}(\Omega)\cos(\Omega t) + G_{B/F}(\Omega)\sin(\Omega t) \right] \quad (12)$$

where $\eta_B$ and $\eta_F$ are the transmission factors of the optical path (in pratice, $\eta_B \sim \eta_F$) and:

$$F_{B/F}(\Omega) = \frac{1}{\Omega} \int_{\theta_{min}}^{\theta_{max}} f(\theta,\Omega)[1 + |\cos\theta|]\sin\theta \, d\theta \quad (13)$$

$$G_{B/F}(\Omega) = \frac{1}{\Omega} \int_{\theta_{min}}^{\theta_{max}} g(\theta,\Omega)[1 + |\cos\theta|]\sin\theta \, d\theta$$

If one assumes that the collection solid angle is $2\pi$ in both experimental configurations, $\theta_{min} = 0$ (resp. $\pi/2$) and $\theta_{max} = \pi/2$ (resp. $\pi$) should be used for the forward (resp. backward) configuration (see Fig. 1). Note that $F_{B/F}(\Omega)$ is in phase with the modulation of the heating and $G_{B/F}(\Omega)$ is in quadrature. Finally, demodulation of the signal power by the lock-in amplifier leads to the PHI signal which magnitude is simply proportional to $P_{B/F}(\Omega)$:

$$P_{B/F}(\Omega) = \eta_{B/F} \sqrt{\alpha_{B/F}} P_i \left[ 2\pi\sqrt{2}.n \frac{\partial n}{\partial T} \frac{P_{abs}}{C\lambda^2 w} \right] \sqrt{F_{B/F}(\Omega)^2 + G_{B/F}(\Omega)^2} \quad (14)$$

## III Experimental results and characterization of the signals

In the following, the experimental details will be given. A schematic of the setup is presented in Figure 2. A non resonant probe beam (632.8 *nm*, HeNe, or single frequency Ti:Sa laser) and an absorbed heating beam (532 *nm*, frequency doubled Nd:YAG laser or tunable cw dye laser) are overlaid and focused on the sample by means of a high NA microscope objective (100x, NA=1.4). The intensity of the heating beam is modulated at Ω by an acousto-optic modulator.

As mentioned previously, the PHI signal can be detected using two different configurations. In the case of the detection of the backward signal, a combination of a polarizing cube and a quarter wave plate is used to extract the interfering probe-reflected and backward-scattered fields. In order to detect the forward signal, a second microscope objective (80x, NA=0.8) is employed to efficiently collect the interfering probe-transmitted and forward-scattered fields. The intensity of the heating beam sent on the NPs ranges from less than $1\,kW/cm^2$ to $\sim 5\,MW/cm^2$ (depending on the desired signal-to-noise ratio and the NP size to be imaged). Backward or forward interfering fields are collected on fast photodiodes and fed into a lock-in amplifier in order to extract the beat signal at Ω. Integration time of 10 *ms* are typically used. Images are formed by moving the sample over the fixed laser spots by means of a 2D piezo-scanner. The size distributions of the gold NPs were checked by transmission electron microscopy (data not shown) and are in agreement with the manufacturer's specification (typically, 10% dispersion in size). The samples were prepared by spin-coating a solution of gold NPs diluted into a polyvinyl-alcohol matrix, (~1% mass PVA) onto clean microscope cover slips. The dilution and spinning speed were chosen such that the final density of NPs in the sample was less than 1 $\mu m^{-2}$. Application of silicon oil on the sample insures homogeneity of the heat diffusion. The index of refraction of that fluid and its thermal conductivity are close to those of common glasses. Thus, there is no sharp discontinuity neither for the thermal parameters nor for the refractive indices at the glass-silicon oil interface and we can consider that the NPs are embedded in a homogeneous medium.

Figure 3 shows a two dimensional PHI image of individual 10 *nm* gold NPs corresponding to the backward (Fig. 3a) and forward (Fig. 3b) signals. Both images show no background signal from the substrate, indicating that the signal arises from the only absorbing objects in the sample, namely the gold aggregates. In both cases the heating intensities were the same (~500 *kW/cm²*) and the NPs are detected with high signal-to-noise ratio (SNR) greater than one hundred.

According to Eq. (14), the resolution of the PHI method depends on the probe and heating beam profiles and also on the dielectric susceptibility profile. Since the spatial extension of the latter is much smaller than the size of the probe beam, the transverse resolution is simply given by the product of the two beams profiles. In figure 4, we study the resolution by imaging a single gold NP with two sets of beam sizes. In the first case, we used low aperture beams with Gaussian profiles measured at the objective focal plane (Fig. 4(a)). In the second case, higher aperture beams are used and their profiles contain diffraction rings arising from the limited aperture of the microscope objective (Fig. 4(b)). In both cases, the transverse profile of the PHI signal is in very good agreement with the product of the two beams profiles. In the first case (Fig.4(a)), the full-width-at-half-maximum (FWHM) of the PHI profile is equal to 365 ± 5nm and reduces to 235 ± 5nm in the second case (Fig.4(b)), in accordance with the products of the beams profiles which are respectively equal to 360 ± 5nm (FWHM) and 213 ± 5nm (FWHM).

The linearity of the PHI signals with respect to $I_{heat}$ was checked on individual 10 *nm* gold NPs (see Fig. 5(a)). Even at high intensities $\left(I_{heat} > 10 \ MW/cm^2\right)$, the PHI signal shows no saturation behavior, rather it is accompanied by fluctuations in the signal amplitude and eventually irreversible damage on the particle[19, 20]. We further confirmed that the peaks stem from single particles by generating the histogram of the signal height for 321 imaged peaks (see Fig. 5(b)). We find a monomodal distribution with a dispersion of 30 % around the mean signal. At a given heating wavelength $\lambda_{heat}$, the PHI signal is proportional to the absorption cross-section $\sigma$. According to the dipolar approximation of Mie theory[21], for small metallic NP with radius $a < \frac{\lambda_{heat}}{2\pi n}$ it scales as the NP's volume. Therefore, our measurements are in good agreement with the spread of 10 % in particle size as evaluated with TEM measurements. The unimodal distribution of the signal values and its dispersion confirm that individual NPs are imaged. As shown in Fig. 3, PHI allows for imaging small gold NPs in both configurations with unprecedented SNR. Owing to this great sensitivity it is possible to detect by optical means and in a far-field configuration gold NPs with diameter down to 1.4 *nm* at moderate intensities (~ 5 *MW/cm²*) and with a SNR > 10 [13].

To clearly demonstrate the linearity of the PHI signals with respect to the volume of the NPs, the size dependence of the absorption cross section of gold NPs (at 532 *nm*, close to the maximum of the SPR) was studied for NP diameters ranging from 1.4 *nm* to 75 *nm*. As expected, a good agreement with a third-order law of the absorption cross-section vs the radius of the particles was found[13].

## IV Frequency dependence

We will now compare the amplitude of PHI backward and forward signals and discuss their dependences with respect to the modulation frequency.

Figures 6 (a-b) present the theoretical dependence of the normalized signal magnitude $\frac{P_B(\Omega)}{\sqrt{\alpha_B P_i}}$ and $\frac{P_F(\Omega)}{\sqrt{\alpha_F P_i}}$ on $\Omega$. Those quantities are proportional to the amplitude of the scattered field in the backward and forward directions ("efficiency" of the incident field scattering). As well, the in-phase and out-of-phase components of the normalized signals are plotted for each case, they are proportional to $F_{B/F}(\Omega)$ and $G_{B/F}(\Omega)$ respectively (Eq. 12). In the calculation, we considered a 10 $nm$ gold NP absorbing $P_{abs}$ = 500 $nW$, and we used $\partial n/\partial T \sim 10^{-4}\ K^{-1}$, $C \sim 2.10^6 J\ K^{-1} m^{-3}, D = 1.8.10^{-8} m^2 s^{-1}, w = 520\ nm$ in Eq. 14.

Two main features can be seen from Figure 6: first, and as expected in any thermal process, the PHI signals exhibit a low-pass behaviour. Second, a clear difference between the forward and the backward normalized signal magnitudes is found at low frequencies, whereas both signals decrease identically in the high frequency domain. Qualitatively, one can understand these features by considering the susceptibility profile as a scattering object with a characteristic size $r_{th}$.

We introduce the size parameter $\xi = \frac{2\pi n r_{th}}{\lambda}$ of the scattering theory[22] and the cut-off frequency $\Omega_C = 2D\left(\frac{2\pi n}{\lambda}\right)^2$ corresponding to $\xi = 1$.

On the one hand, for high frequencies ($\Omega \gg \Omega_C$) when the scattering object is small compared to the incident probe beam wavelength ($\xi \ll 1$), the angular distribution of the scattered light is symmetric with respect to the focal plane[22] leading to identical backward and forward PHI signals. Furthermore, the in-phase components $F_{B/F}(\Omega)$ decrease as $1/\Omega^2$ and the out-of-phase component $G_{B/F}(\Omega)$ decrease as $1/\Omega$. This is the typical response of a driven system in a dissipative medium. As a consequence, one finds that $\frac{P_B(\Omega)}{\sqrt{\alpha_B P_i}}$ and $\frac{P_F(\Omega)}{\sqrt{\alpha_F P_i}}$ are identical in the high frequency domain and decrease as $1/\Omega$.

On the other hand, for low frequencies ($\Omega \leq \Omega_C, \xi \geq 1$), the scattering object is large enough so that forward scattering becomes significantly more efficient than backward scattering[22], as shown in figure 6 where $\frac{P_F(\Omega)}{\sqrt{\alpha_F P_i}}$ is greater than $\frac{P_B(\Omega)}{\sqrt{\alpha_B P_i}}$.

Fig. 6 (a-b) also present experimental data points from measurements performed on a single 10 *nm* gold NP. The modulation frequency ranged from 100 kHz to 15 MHz. As mentioned above, experimental data were adjusted by the model resulting from Eq.(14) with *D* as the only adjustable parameter. For both configurations, limitations of the collection efficiency by the objective lenses were neglected. A good quantitative agreement between the experimental data and the theoretical forms of $\frac{P_F(\Omega)}{\sqrt{\alpha_F P_i}}$ and $\frac{P_B(\Omega)}{\sqrt{\alpha_B P_i}}$ is obtained for the same value of $D = (1.8 \pm 0.1) \, 10^{-8} \, m^2 s^{-1}$ in both configurations. This value is in good agreement with the technical specifications of the embedding medium. We also reported the values of the typical sizes of the scattering objects $r_{th}$ mentioned above and reported them on the top axis of Fig. 6 (a-b).

To further test our theoretical description of the PHI signals, Eq. (14) can be used to estimate numerically the PHI power detected at the detector in both configurations. A single 10 *nm* gold having an absorption cross section of $\sigma \sim 5.10^{-13} \, cm^2$ at 532 *nm* and absorbing $P_{abs} \sim 200 \, nW$ when illuminated by a laser intensity of $I_{Heat} \sim 400 \, kW/cm^2$, one expects for a incident probe power $P_i = 7.5 \, mW$, a frequency $\Omega/2\pi = 700 \, kHz$ and a transmitted (resp.

reflected) power $P_{LO} = \alpha_F P_i \approx P_i$ (resp. $P_{LO} = \alpha_B P_i \approx 15\mu W$), to detect a PHI power of $P_B \sim 10 nW$ and $P_F \sim 450 nW$ in the backward and forward directions. Experimentally and after calibration of the detection chain, we measured PHI powers of $P_B \sim 4 nW$ and $P_F \sim 60 nW$ in the backward and forward directions (mean of 80 signals from different individual 10 nm gold NPs). A qualitative agreement is found between the modeling and the experimental results. The discrepancy is certainly due to the assumption of a spherical reference wave in the calculation of the beatnote power. This optimal case is not reached in our experiments even with high NA microscope objectives.

In any case, the forward PHI signal is significantly larger than the backward one, due to a more efficient heterodyne amplification in the forward direction ($\alpha_F >> \alpha_B$). However, in the case of shot noise limited detection, the two configurations are expected to give identical SNRs, except at frequencies lower than $\Omega_c$ for which the transmission scheme is expected to give better SNRs as forward scattering becomes significantly more efficient than backward scattering. Experimentally, shot noise limited detection is difficult to obtain for low frequencies, and it is only in the forward detection scheme and for sufficiently high incident probe powers $(P_i > 1 mW)$ that it is achieved. Nevertheless the excess noise was low enough to obtain close-to-optimal SNR in the backward detection scheme (see Fig.3).

## V Absorption spectroscopy of individual metallic NPs

PHI not only allows for highly sensitive detection of gold NPs but also opens new pathways towards absorption spectroscopy of single metallic NP and more generally of non-luminescent nano-objects. Several theoretical models predicted the existence of so-called intrinsic size effects in the optical response of metallic NPs with sizes significantly smaller than the electron mean free path[21, 23]. Limitation of the electron mean free path as well as chemical interface damping[21, 24] increase the damping rate of the Surface Plasmon Resonance (SPR), thus leading in the time domain to shorter dephasing times (down to a few fs). Because of such intrinsic size effects, the dielectric constant of the NPs differs from that of the bulk value and must include an additional surface damping contribution. Experimental studies on ensembles of metallic NPs revealed the existence of such effects [21, 25]. However a quantitative description of those effects was made difficult mostly because of inhomogeneous broadening. In order to overcome this shortcoming, PHI was used to record absorption spectra of individual gold NPs with diameters down to 5 $nm$[14]. Figure 7 represents the absorption spectra of 2 single gold NPs with diameters of 33 $nm$ and 5 $nm$ respectively, performed in the wavelength range 515-580 $nm$ (i.e. photon energy range 2.41-2.14 $eV$). The values of peak resonance energies are not particularly affected by intrinsic size effects. On the contrary, a significant increase in width of the resonance $\Gamma$ is clearly visible that cannot be described by Mie theory using the bulk values of the gold dielectric constant[26]. We found a good agreement between the experimental widths and Mie simulations by introducing the size dependent correction to the bulk dielectric constant that accounts for the presence of intrinsic size effects in the absorption spectra of small gold NPs[14]. Although the existence of intrinsic size effects in the optical response of gold NPs was unambiguously revealed, part of the damping processes are due to interband transition, which makes it difficult to connect the widths of the Plasmon resonances to the damping rate. Indeed, for gold, the energy threshold for interband transitions lies at ~2.4 $eV$ and is preceded by an absorption tail starting at about 1.8 $eV$. Consequently, the SPR spectra are asymmetric and defining a full-width-at-half maximum of absorption spectra for gold NPs is delicate, and even impossible for very small particles. A way to circumvent this additional damping would consist in red-shifting resonant energies towards photon energy smaller than the onset of interband transition. This can be done either by embedding spherical gold NPs in a matrix with high refractive index or by studying the SPR of the long axis mode in gold nanorods [6, 27] or even by using core-shell NP[28].

Another possibility consists in using spherical silver NPs instead of gold NPs[29]. Indeed for silver, the resonant energy of spherical NPs embedded in a medium with refractive index $n = 1.5$ is at ~3 $eV$, whereas interband transitions start at ~3.7 $eV$[26]. Thus silver NPs are well adapted for studying the size dependence of ultrafast dephasing of the SPR at the single particle level. A limitation of silver NP is however their weak photostability due to photo-oxydation[30]. This limitation can be surpassed by using encapsulated silver NPs[31] e.g. with PEG (poly(ethylene glycol))[32]. Figure 8(a) shows a three dimensional representation of PHI image of individual PEG-coated silver NPs with average diameter of 5.3 $nm$. For this image, a modulated diode laser (Coherent Compass) emitting at 405 $nm$ was used as heating laser with heating intensity of only $\sim 50\, kW/cm^2$. As expected, we found that the SNR obtained with silver NPs is about 10 times higher than that of gold NPs of the same size, when identical heating intensities are used to excite the NPs at the peak of their SPR resonance. The PHI signal measured on PEG-coated silver NPs was stable during more than one minute (see Fig. 8(b)). For those silver nanoparticles, the size distribution checked by TEM (data not shown) was poor: 20%. This translates in an dispersion of 90% in the distribution of the measured PHI signals (653 imaged peaks (see Fig. 8(b)). For such experiments, sensitivity at the single particle level is all the more necessary as the heterogeneity in size and shape is obviously larger for silver NPs than for gold NPs. In addition, expected progresses in the synthesis of silver NPs such that narrower size distributions NPs can be obtain should permit a quantitative study of the SPR of individual silver NPs much smaller than 5 $nm$ in diameter.

# VI. Conclusion and Perspectives

In conclusion, we have presented the theoretical and experimental characterizations of the Photothermal Heterodyne Imaging technique which allows for the detection of individual non fluorescent nano-objects such as metallic nanoparticles (gold or silver) or even semiconductor nanocrystals[13]. The photothermal heterodyne signal was derived through a model based on classical scattering theory. Experiments on individual gold nanoparticles provided results in good agreement with our calculations. Photothermal Heterodyne Imaging, which is not limited by most shortcomings of fluorescence microscopy such as scattering background or photobleaching allows for long observation times with high signal to noise ratios.

The high sensitivity provided by PHI, coupled to a better control of the NPs synthesis should allow original studies of light-NP interaction processes and we foresee diverse applications in different fields such as in plasmonics or bioimaging. For instance, gold NPs have been used as labels for biological applications [8,33]. In this context, PHI is potentially the most sensitive method to detect gold NPs in live cells by optical means. For such applications, the local increase of temperature in the vicinity of the NP must not perturb the biological function of the specimens under study. Since the thermal conductivity of metals is much higher than that of the surrounding medium, the temperature inside a spherical NP can be considered as uniform and equal to the temperature at its surface. This temperature writes: $T_{surf} = \frac{P_{abs}}{4\pi\kappa a}$ and decreases as the inverse of the distance from the NP surface. For 5 *nm* gold NPs in aqueous medium, a heating intensity of 500 *kW/cm²* leads to a surface temperature of ~ 1.5 *K*, sufficiently low for live cell integrities.

# Acknowledgments

We wish to thank D. Ferning and C. Brust for the synthesis of silver nanoparticles and Philippe Tamarat for helpful discussions. This research was funded by CNRS and the French Ministry for Education and Research (ACI Nanoscience and DRAB) and the Région Aquitaine.

# Figure captions

## Figure 1

(Color online) Relative position of the observation point M to the nanoparticle and probe beam. The interaction of an intensity modulated heating beam (not represented) with a single nanoparticle (black dot) leads to a photothermally induced index of refraction profile (gray area). Propagation of the probe beam through this profile, produces a frequency shifted scattered field which interferes with either the reflected or the transmitted probe field, leading to a photothermal heterodyne signal.

## Figure 2

(Color online) Experimental setup showing the detection schemes of PHI signal in the backward and forward directions.

## Figure 3

(Color online) 2D representations of two photothermal heterodyne images of the same area (6x6 $\mu m^2$) containing individual 10 $nm$ gold NPs, taken in the backward (a) and forward directions (b). Scale bars are 1 $\mu m$.

## Figure 4

Transverse resolution of the PHI method with low (a) and high (b) aperture of the beams. The measured profiles of the probe beam (dashed-dotted lines) and heating beam (dashed lines) are indicated. In both cases, the profile of the PHI signal from a single 10nm gold NP (circles) is in very good agreement with the product of the two beams profiles (solid line). The transverse resolutions of the PHI signals are (a) 365 ±5 *nm* (FWHM) and (b) 235±5 *nm* (FWHM).

## Figure 5

(a) Signal obtained from an individual 10 nm gold nanoparticle (circles) as a function of the heating intensity. The data points are adjusted by a linear fit (solid line). (b) Signal histogram of 321 peaks detected in a sample prepared with 10nm gold NPs. The monomodal shape of the distribution reveals that individual NPs are detected.

## Figure 6

Theoretical (solid black lines) and experimental (circles) dependence of the (a) forward and (b) backward signals. $\frac{P_F(\Omega)}{\sqrt{\alpha_F P_i}}$ and $\frac{P_B(\Omega)}{\sqrt{\alpha_B P_i}}$ as a function of the modulation frequency $\Omega$. The in phase (dashed gray line) and out of phase (dotted gray line) components as well as the values of $r_{th}$ are also indicated. Measurements were performed in both configurations on a single 10 nm gold NP.

## Figure 7

Normalized absorption spectra of 2 single gold NPs of respective diameters equal to 33 *nm* (open circles), and 5 *nm* (open squares). The extracted red width at half maximum $\Gamma$ is shown on the both NP spectra. The experimental values are compared with simulations based on Mie theory (solid lines) using size dependant modification in the dielectric constant of gold.

**Figure** 8

(Color online) (a) 3D representation of a photothermal heterodyne image (5×5 μm$^2$) containing individual 5 *nm* silver NPs. (b) Temporal trace of the PHI signal arising from a single 5 *nm* silver NP. (c) Signal histogram of 653 peaks detected in a sample containing 5 *nm* silver NPs.

Figure 1 (color online)

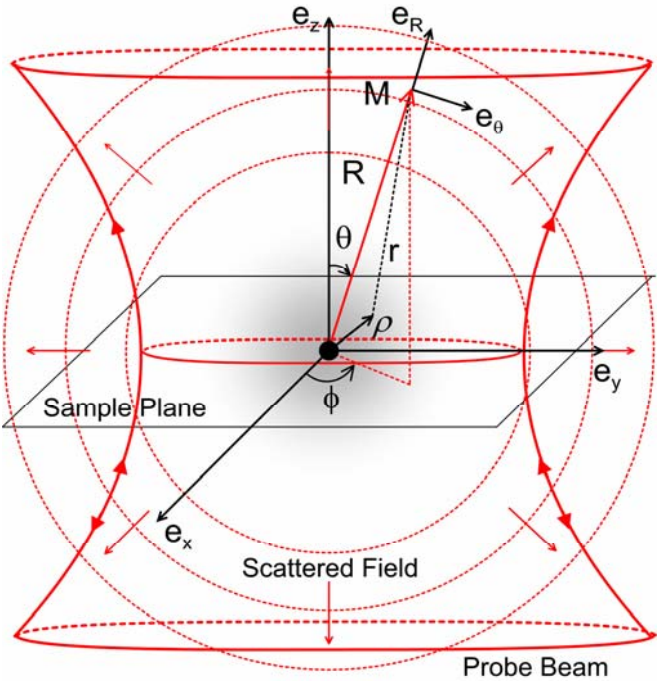

Figure 2 (color online)

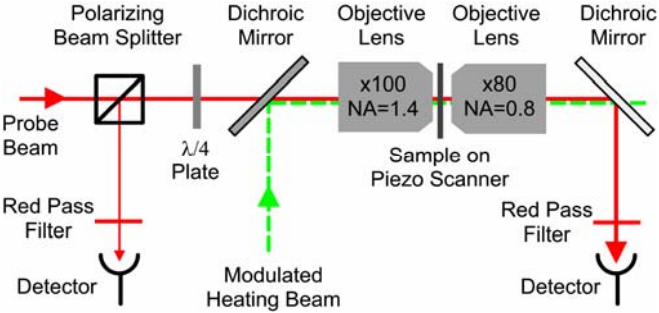

Figure 3 (color online)

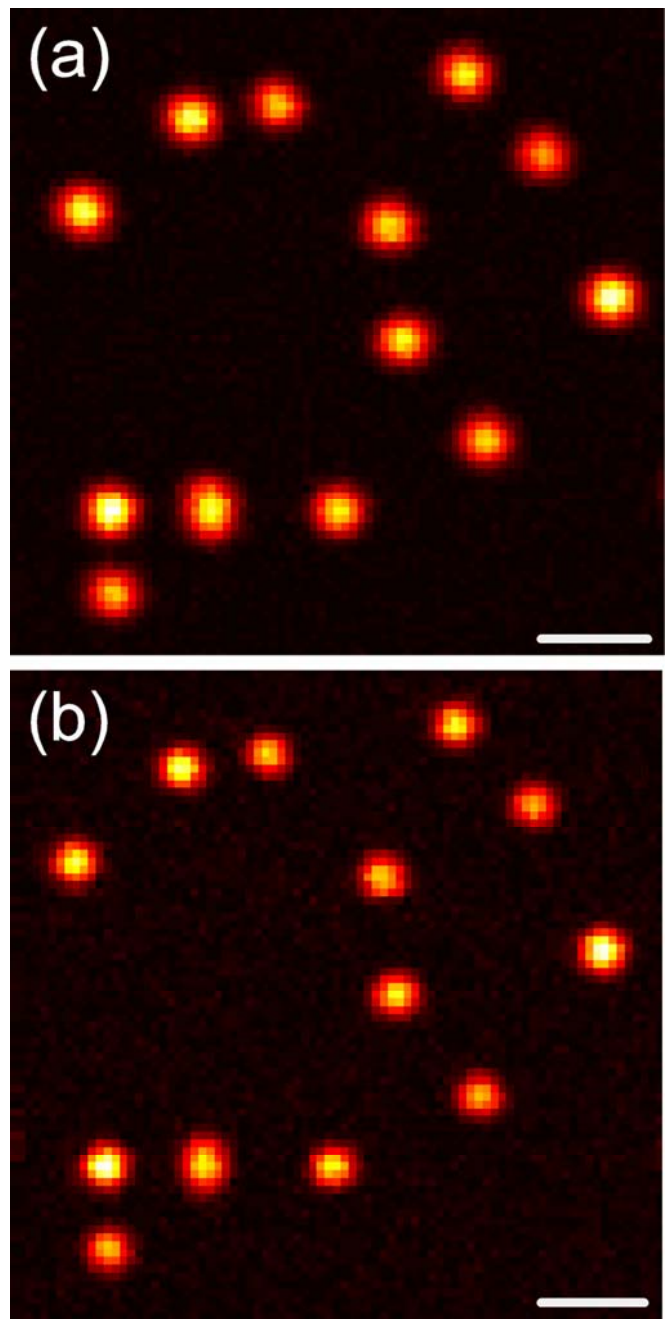

Figure 4

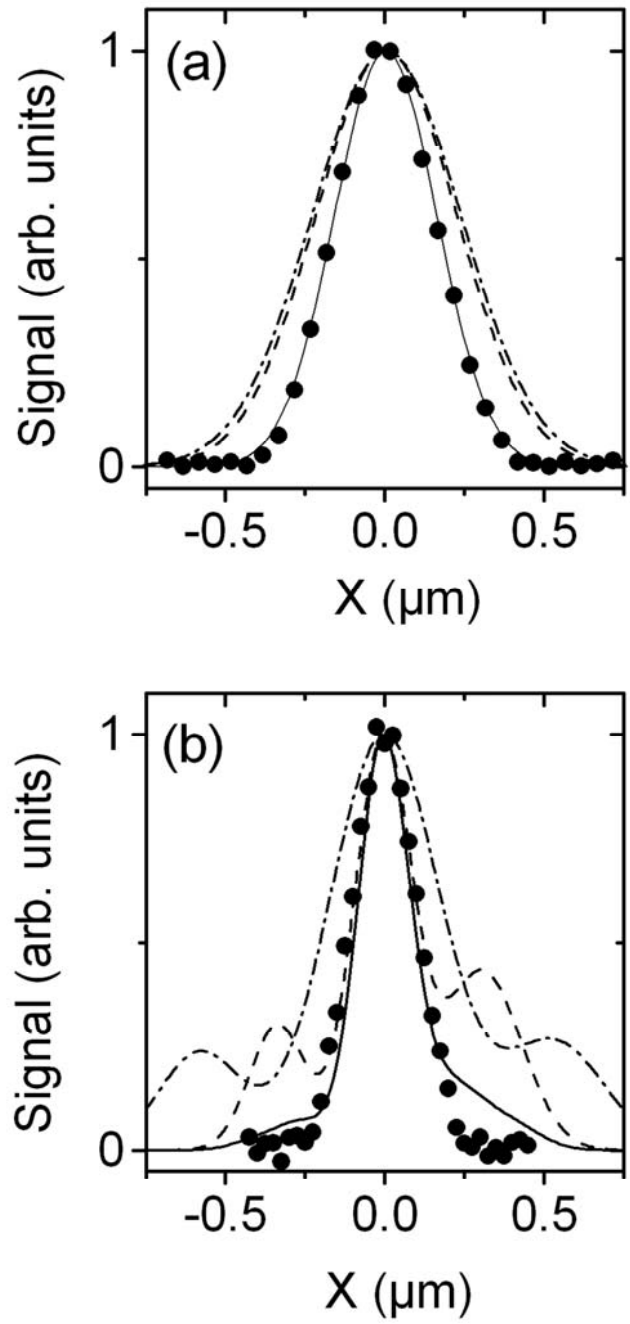

Figure 5

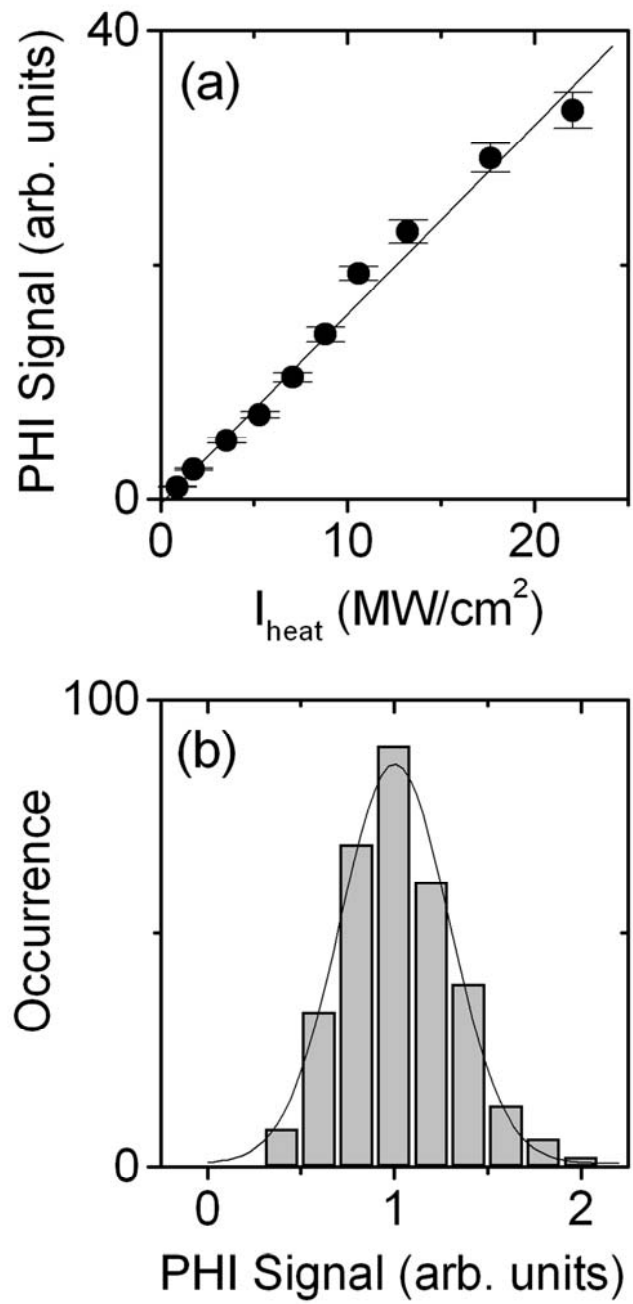

Figure 6

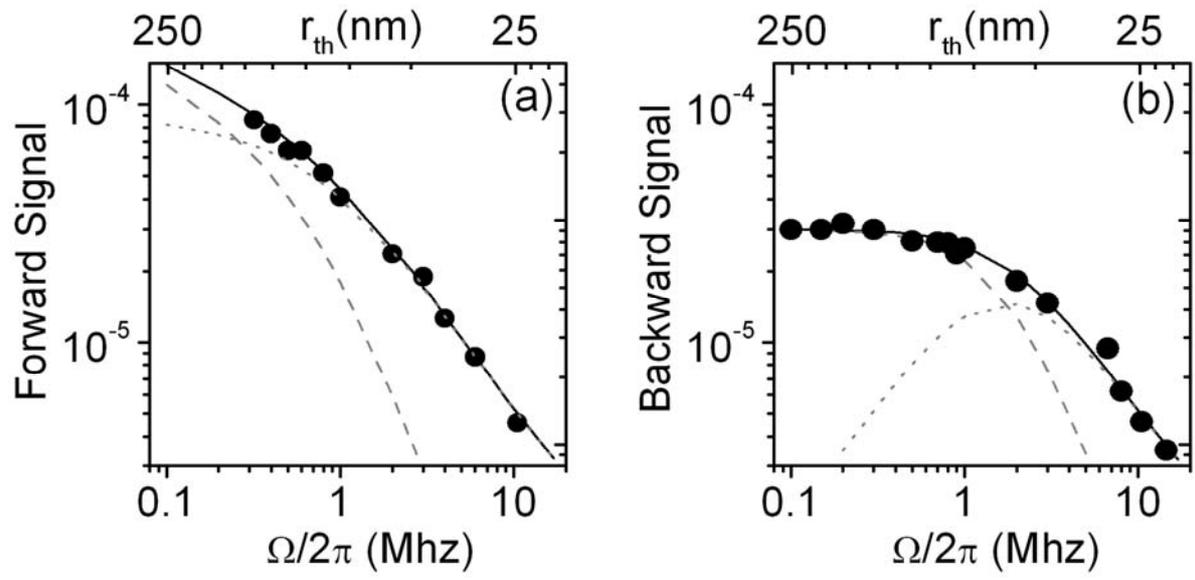

Figure 7

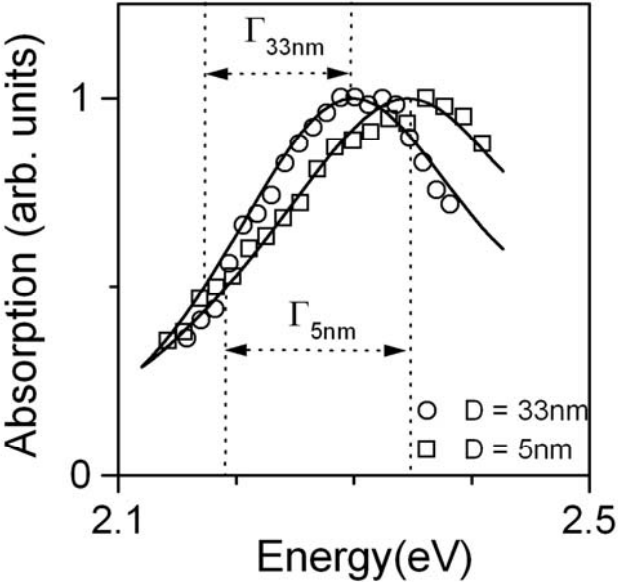

Figure 8 (color online)

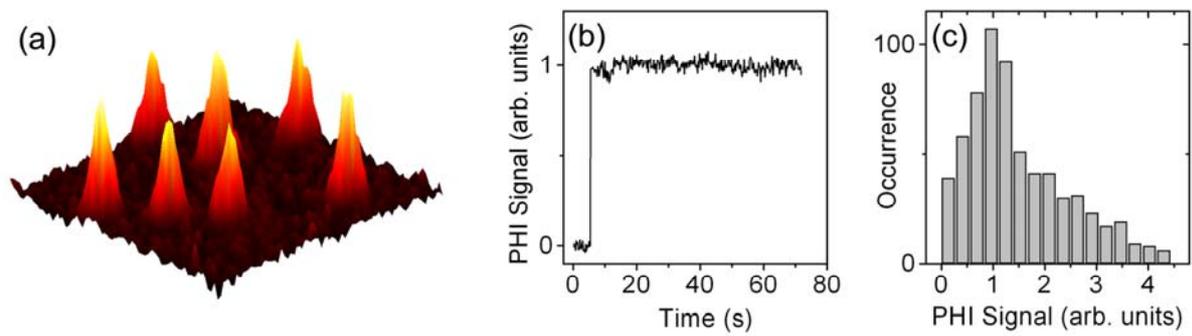